\documentstyle[12pt,graphicx,pstricks,braket,amsmath,amssymb,epstopdf,pdfpages,cite,tabu,hyperref]{article}
\numberwithin{equation}{section}
\begin{document}

\def\AEF{A.E. Faraggi}

\def\JHEP#1#2#3{{JHEP} {\textbf #1}, (#2) #3}
\def\vol#1#2#3{{\bf {#1}} ({#2}) {#3}}
\def\NPB#1#2#3{{\it Nucl.\ Phys.}\/ {\bf B#1} (#2) #3}
\def\PLB#1#2#3{{\it Phys.\ Lett.}\/ {\bf B#1} (#2) #3}
\def\PRD#1#2#3{{\it Phys.\ Rev.}\/ {\bf D#1} (#2) #3}
\def\PRL#1#2#3{{\it Phys.\ Rev.\ Lett.}\/ {\bf #1} (#2) #3}
\def\PRT#1#2#3{{\it Phys.\ Rep.}\/ {\bf#1} (#2) #3}
\def\MODA#1#2#3{{\it Mod.\ Phys.\ Lett.}\/ {\bf A#1} (#2) #3}
\def\RMP#1#2#3{{\it Rev.\ Mod.\ Phys.}\/ {\bf #1} (#2) #3}
\def\IJMP#1#2#3{{\it Int.\ J.\ Mod.\ Phys.}\/ {\bf A#1} (#2) #3}
\def\nuvc#1#2#3{{\it Nuovo Cimento}\/ {\bf #1A} (#2) #3}
\def\RPP#1#2#3{{\it Rept.\ Prog.\ Phys.}\/ {\bf #1} (#2) #3}
\def\APJ#1#2#3{{\it Astrophys.\ J.}\/ {\bf #1} (#2) #3}
\def\APP#1#2#3{{\it Astropart.\ Phys.}\/ {\bf #1} (#2) #3}
\def\EJP#1#2#3{{\it Eur.\ Phys.\ Jour.}\/ {\bf C#1} (#2) #3}
\def\etal{{\it et al\/}}
\def\notE6{{$SO(10)\times U(1)_{\zeta}\not\subset E_6$}}
\def\E6{{$SO(10)\times U(1)_{\zeta}\subset E_6$}}
\def\highgg{{$SU(3)_C\times SU(2)_L \times SU(2)_R \times U(1)_C \times U(1)_{\zeta}$}}
\def\highSO10{{$SU(3)_C\times SU(2)_L \times SU(2)_R \times U(1)_C$}}
\def\lowgg{{$SU(3)_C\times SU(2)_L \times U(1)_Y \times U(1)_{Z^\prime}$}}
\def\SMgg{{$SU(3)_C\times SU(2)_L \times U(1)_Y$}}
\def\Uzprime{{$U(1)_{Z^\prime}$}}
\def\Uzeta{{$U(1)_{\zeta}$}}

\newcommand{\cc}[2]{c{#1\atopwithdelims[]#2}}
\newcommand{\bev}{\begin{verbatim}}
\newcommand{\beq}{\begin{equation}}

\newcommand{\beqa}{\begin{eqnarray}}
\newcommand{\beqn}{\begin{eqnarray}}
\newcommand{\eeqn}{\end{eqnarray}}
\newcommand{\eeqa}{\end{eqnarray}}
\newcommand{\eeq}{\end{equation}}
\newcommand{\beqt}{\begin{equation*}}
\newcommand{\eeqt}{\end{equation*}}
\newcommand{\Eev}{\end{verbatim}}
\newcommand{\bec}{\begin{center}}
\newcommand{\eec}{\end{center}}
\newcommand{\bes}{\begin{split}}
\newcommand{\ees}{\end{split}}
\def\ie{{\it i.e.~}}
\def\eg{{\it e.g.~}}
\def\half{{\textstyle{1\over 2}}}
\def\nicefrac#1#2{\hbox{${#1\over #2}$}}
\def\third{{\textstyle {1\over3}}}
\def\quarter{{\textstyle {1\over4}}}
\def\m{{\tt -}}
\def\mass{M_{l^+ l^-}}
\def\p{{\tt +}}

\def\slash#1{#1\hskip-6pt/\hskip6pt}
\def\slk{\slash{k}}
\def\GeV{\,{\rm GeV}}
\def\TeV{\,{\rm TeV}}
\def\y{\,{\rm y}}

\def\l{\langle}
\def\r{\rangle}
\def\LRS{LRS  }

\begin{titlepage}
\samepage{
\setcounter{page}{1}
\rightline{LTH--974} 
%\rightline{arXiv:????.????}
\vspace{1.5cm}

\begin{center}
 {\Large \bf 
 Proton Stability,\\Gauge Coupling Unification\\ \medskip and a Light $Z^\prime$ in Heterotic--string Models}
\end{center}

\begin{center}
%\vspace{1.cm}
Alon E. Faraggi\footnote{
		                  E-mail address: faraggi@amtp.liv.ac.uk}
and 
Viraf M. Mehta\footnote{ E-mail address:
	                          Viraf.Mehta@liv.ac.uk}
\\
\vspace{.25cm}
{\it Department of Mathematical Sciences\\
University of Liverpool, Liverpool, L69 7ZL, United Kingdom}
\end{center}

\begin{abstract}

We explore the phenomenological viability of a light
$Z^\prime$ in heterotic--string models, whose 
existence has been motivated by proton stability arguments. 
A class of quasi--realistic 
string models that produce such a viable $Z^\prime$  are the Left--Right
Symmetric (LRS) heterotic--string models in the free fermionic formulation. 
A key feature of these models is that the matter charges under
$U(1)_{Z^\prime}$ do not 
admit an $E_6$ embedding. The light $Z^\prime$ in the LRS heterotic--string 
models forbids baryon number violating operators, while 
allowing lepton number violating operators, hence suppressing 
proton decay yet allowing for sufficiently small neutrino masses 
via a seesaw mechanism. We show that the constraints imposed 
by the gauge coupling data and heterotic--string coupling unification
nullify the viability of a light $Z^\prime$ in these  models. We further
argue that agreement with the gauge coupling data 
necessitates that the $U(1)_{Z^\prime}$ charges admit an $E_6$ embedding. 
We discuss how viable string models with this property may be constructed.

\end{abstract}
\smallskip}
\end{titlepage}

\section{Introduction}

The discovery of the Higgs boson at the LHC lends further 
credence to the hypothesis that the Standard Model (SM) provides 
a viable effective parametrisation of all subatomic interactions
up to the GUT or heterotic--string unification scales.
Support for this possibility stems from: the 
matter gauge charges; proton longevity; suppression of
neutrino masses; and the logarithmic evolution of the 
SM parameters in its gauge and matter sectors.
Preservation of the logarithmic running in the SM
scalar sector entails that it must be augmented by a new symmetry. 
A concrete framework that fulfils the task is given by supersymmetry.

The supersymmetric extension of the SM introduces
dimension four and five baryon and lepton number violating operators
that mediate proton decay. This problem is particularly
acute in the context of heterotic--string derived constructions, 
in which one cannot assume the existence of global or local
discrete symmetries that simply forbid the undesired operators.
%%%%%%%%%%%%%%%%%%%%%%%%%%%%%
Indeed, the issue has been examined in the past by a number 
of authors \cite{discrete}. 
The avenues explored range from the existence of matter 
parity at special points in the moduli space of specific
models, to the emergence of non--abelian custodial symmetries
in some compactifications. However, a caveat to these arguments
is that in addition to suppressing the proton decay mediating 
operators, one must also ensure that the mass terms 
of left--handed neutrinos are sufficiently suppressed.
That is, while baryon number should be conserved to ensure
proton longevity, lepton number must be broken to allow
for suppression of left--handed neutrino masses.
In heterotic--string constructions, due to the absence of
higher--order representations of the Grand Unified Theory
\cite{dmr}, one typically has to break lepton number
by one unit, which generically results in both
lepton and baryon number violation. 
%%%%%%%%%%%%%%%%%%%%%%%%%%%%%%%%
%A viable solution to this conundrum is obtained if 
An alternative solution to this conundrum is obtained if 
an additional $U(1)$ gauge symmetry, beyond the SM 
gauge group, remains unbroken down to low scales.
An additional abelian gauge symmetry, which is 
broken near the TeV scale, may also explain the 
suppression of the $\mu$--term in the supersymmetric
potential \cite{cleavermuterm}.

The possibility of a low scale $Z^\prime$ 
arising from heterotic--string inspired models 
has a long history
and continues to attract wide interest \cite{zphistory}.
Surprisingly, however, keeping a $Z^\prime$ in 
explicit string derived constructions, unbroken down
to the low scale, turns out to be notoriously 
difficult, as such an extra symmetry 
must satisfy a variety of phenomenological constraints. 
Obviously, to play a role in the suppression of 
proton decay mediating operators (PDMOs) implies that the 
SM matter states are charged under this symmetry. 
While forbidding baryon number violation, it should allow for
lepton number violation, required for the suppression of 
neutrino masses. 
Furthermore, it should be family universal, otherwise 
there is a danger of generating Flavour Changing Neutral 
Currents (FCNC), or of generating the PDMOs via mixing. The additional symmetry should also 
allow for the fermion Yukawa couplings to electroweak Higgs doublets and must 
be anomaly free. 
Explicit string models that do give rise to an extra $U(1)$ 
symmetry with the required properties are the left--right symmetric 
models of \cite{lrs,ps}. 
The existence of the required symmetry in explicit string constructions 
ensures that, in these examples, the extra $U(1)$ is free of any 
gauge and gravitational anomalies. In \cite{fm1} we
constructed toy string--inspired models that are 
compatible with the charge assignments in the string
derived models. 
In these models, the proton lifeguarding
extra $U(1)$ symmetry can, in principle, remain unbroken 
down to low scales. 

An additional constraint that must be imposed on the extra 
gauge and matter states that arise in the $Z^\prime$ models,
is compatibility with the gauge parameters, $\sin^2\theta_W(M_Z)$ 
and $\alpha_3(M_Z)$. The perturbative
heterotic--string predicts that all the gauge couplings are unified 
at the string unification scale, $M_S$, which is of the order $5\cdot 10^{17}{\rm GeV}$. 
Nonperturbatively, the heterotic--string can be pushed
to the GUT unification scale, $M_{\mathrm{GUT}}$, of the order $2\cdot 10^{16}{\rm GeV}$ 
\cite{witten}.
In this paper we study the constraints that are imposed on the 
string inspired $Z^\prime$ models by gauge coupling unification and 
show that the gauge coupling data is not in agreement with 
the left--right symmetric heterotic--string models. The origin for the disagreement
lies in the specific $U(1)_{Z^\prime}$ charges, which do not 
admit an $E_6$ embedding. For comparison we also perform the 
analysis for $U(1)_{Z^\prime}$ charges that maintain the 
$E_6$ embedding and show that, in this case, agreement with 
the data is achieved. We discuss how viable 
string derived models that preserve the $E_6$ embedding may be 
constructed. 

\section{Additional $U(1)$s in free fermionic models}\label{review}

In this section we review the structure of the free fermionic 
models. We focus on the extra $U(1)$ symmetries 
that arise in the models and the charges of the matter states. 
We elaborate on the gauge symmetry breaking patterns induced
by the Generalised GSO (GGSO) projections but concentrate here on
the group theory structure and the matter charges. Further 
details of the free fermionic models and 
their construction are found in earlier literature 
\cite{lrs,fff,fsu5,slm, alr,nahe, su421, classification, exophobic}.
The free fermionic models 
correspond to $Z_2\times Z_2$ orbifold compactifications 
at special points in the moduli space \cite{z2xz2}. 
It should be emphasized
that our results are applicable to the wider range of orbifold
models because they merely depend on the symmetry breaking 
patterns of the observable gauge symmetry.

Free fermionic heterotic--string models are constructed by 
specifying a consistent set of boundary condition basis vectors
and the associated one--loop GGSO phases \cite{fff}. 
These basis vectors span a finite additive group, $\Xi$, where the 
physical states of a given sector, $\alpha\in\Xi$, are obtained by
acting on the vacuum with bosonic and fermionic operators and 
by applying the GGSO projections. The $U(1)$ charges, with respect
to the unbroken Cartan generators of the four dimensional gauge 
group, are given by: 
\beq 
Q(f)= {1\over2} \alpha(f) + F(f),
\label{qcharge}
\eeq
where $\alpha(f)$ is the boundary condition of the 
complex world--sheet fermion $f$ in the sector $\alpha$, and 
$F_\alpha(f)$ is a fermion number operator counting each mode of 
$f$ once ($f^*$ minus once). 
For periodic fermions with
$\alpha(f)=1$, the vacuum is a spinor representing the Clifford
algebra of the zero modes. 
For each periodic complex fermion, $f$,
there are two degenerate vacua, ${\vert +\rangle}\mathrm{ and }{\vert -\rangle}$, 
annihilated by the zero modes, $f_0$ and
${f_0^*}$, with fermion numbers  $F(f)=0,-1$ respectively.

Three generation models in the free fermionic construction have been
obtained by using two constructions: the first were the NAHE
based models \cite{nahe}; and the second class of models are 
those constructed by the classification method of \cite{classification}.
The important distinction  between the two cases is that the 
latter has only been applied for symmetric orbifolds, whereas, in the 
former, most of the constructions utilise asymmetric boundary conditions. 

In NAHE based models \cite{lrs,fsu5,slm,alr, su421} 
the first set of five
basis vectors,  $\{{{\bf 1},S,b_1,b_2,b_3}\}$, 
are fixed; $b_1$, $b_2$ and $b_3$ correspond to the 
three twisted sectors of the $Z_2\times Z_2$ orbifold 
and $S$ is the spacetime supersymmetry generator.
The gauge symmetry at the level of the 
NAHE set is $SO(10)\times SO(6)^3\times E_8$ with $N=1$ spacetime
supersymmetry. 
The second stage of the construction consists of adding three
additional basis vectors to the NAHE set.
The additional vectors reduce the number of generations 
to three and simultaneously break the four dimensional 
group. The $SO(10)$ symmetry is broken to one
of its maximal subgroups:
$SU(5)\times U(1)$ (FSU5) \cite{fsu5};
$SU(3)\times SU(2)\times U(1)^2$ (SLM) \cite{slm};
$SO(6)\times SO(4)$ (PS) \cite{alr};
$SU(3)\times U(1)\times SU(2)^2$ (LRS) \cite{lrs}; 
and  $SU(4)\times SU(2)\times U(1)$ (SU421) \cite{su421}. 

An important distinction between the last two cases and the 
first three is in regard to the anomalous $U(1)_A$ symmetry
that arises in these models 
\cite{lrs,  su421,cleaverau1}. The Cartan subalgebra of the observable
rank eight gauge group is generated by eight complex fermions,
denoted by $\{{\bar\psi}^{1,\cdots,5}, {\bar\eta}^{1,2,3}\}$,
where ${\bar\psi}^{1,\cdots,5}$ are the Cartan generators of the
$SO(10)$ group and ${\bar\eta}^{1,2,3}$ generate three
$U(1)$ symmetries, denoted by $U(1)_{1,2,3}$.
In the FSU5, PS and SLM cases the $U(1)_{1,2,3}$, as well as their
linear combination, 
\beq
U(1)_\zeta = U(1)_1+U(1)_2+U(1)_3,
\label{u1zeta}
\eeq
are anomalous, whereas in the LRS and SU421 models they are anomaly free.
The distinction can be seen to arise from the symmetry breaking patterns 
induced in the two cases from the underlying $N=4$ 
toroidal model in four dimensions. Starting from the $E_8\times E_8$, in the first
case the symmetry is broken to $SO(16)\times SO(16)$ by the choice of 
GGSO projection phases in the fermionic models, or equivalently by 
a Wilson line in the corresponding orbifold models. The basis 
vectors $b_1$ and $b_2$ break the symmetry further to 
$SO(10)\times U(1)^3\times SO(16)$. 
Alternatively, we can implement the $b_1$ and $b_2$ twists in the $E_8\times E_8$
vacuum, which break the gauge symmetry to $E_6\times U(1)^2\times E_8$. 
The Wilson line breaking then reduces the symmetry to 
$SO(10)\times U(1)_\zeta\times U(1)^2
\times SO(16)$. It is then clear that the $U(1)_\zeta$ becomes
anomalous because 
of the $E_6$ symmetry breaking to $SO(10)\times U(1)_\zeta$ and the 
projection  of some states from the spectrum by the 
GGSO projections \cite{cleaverau1}. 
On the other hand, the LRS and SU421 heterotic--vacua arise from an $N=4$ vacuum 
with $E_7\times E_7\times SO(16)$ gauge symmetry \cite{lrs,su421}. In this case, one of the $E_7$ 
factors produces the observable gauge symmetry and the second is hidden. 
The important point here is that these models circumvent the $E_6$
embedding. Hence, in these cases, the $U(1)_\zeta$ does not 
have an $E_6$ embedding and therefore remains anomaly free. 

The case of the symmetric orbifolds studied in \cite{classification}
only allows for models with an $E_6$ embedding of $U(1)_\zeta$. Thus, in these
models $U(1)_\zeta$ is, generically, anomalous. There is, however, 
a class of models in which it is anomaly free. This is the case in the self--dual 
models under the spinor--vector duality of \cite{spinvecdual}. 
In these models the number 
of $SO(10)$ spinorial $\mathbf{16}$ representations and the number of vectorial $\mathbf{10}$
representations, arising from the twisted sectors is identical, although the
$E_6$ symmetry is broken. This situation occurs when the spinorial
and vectorial representations are obtained from different fixed
points of the $Z_2\times Z_2$ toroidal orbifold. A self--dual, three generation 
model with unbroken $SO(10)$ symmetry is given in ref. \cite{classification}, 
however, a viable model, of this type,
with broken $SO(10)$ symmetry has not been constructed to date. 

Alternatively, we may construct $U(1)_\zeta\subset E_6$ as an anomaly free
combination by following a different symmetry breaking pattern 
to the $E_6\rightarrow SO(10)\times U(1)$ discussed above. 
Originally, the $E_6\rightarrow SO(10)\times U(1)$ breaking is achieved 
by projecting the vector bosons that 
arise in the spinorial $\mathbf{128}$ representation of $SO(16)$ and 
enhance the $SO(16)$ symmetry to $E_8$. We may construct 
models in which these vector bosons are not projected and, thus, 
the $E_6$ symmetry is broken to a different subgroup. 
Examples of such models include the three generation 
$SU(6)\times SU(2)$ models of \cite{bfgrs}. In this case, the
$U(1)_\zeta$ is anomaly free by virtue of its embedding
in the enhanced symmetry.

\section{Gauge coupling analysis}\label{gca}
In this section we present a comparative analysis of the two classes mentioned above.  
It will be instructive to specify a model in each class:
\begin{itemize}
\item Model I: This model was first presented in \cite{fm1}. 
In this case the extra \Uzeta~ 
does not admit an $E_6$ embedding, {\it i.e.} \notE6.  
\vspace{8pt}
\item Model II: This model preserves the $E_6$ embedding 
of the \Uzeta ~and is akin to $Z^{\prime}$ 
models arising in string inspired $E_6$ models \cite{zphistory}. 
\end{itemize}
Before proceeding with the gauge coupling analysis, it is instructive 
to detail the symmetry breaking patterns applicable to both models. 
The SM gauge group will be embedded, for our
analysis, in $SO(10)$. As previously mentioned, this is broken to the LRS gauge group 
via the addition of basis vectors, 
$\alpha, \beta, \mbox{ and } \gamma$ at the string scale, $M_S$.
The $SU(2)_R$ is then broken at some intermediate
scale, $M_R$.  An anomaly free $U(1)$ combination that remains is 
the \Uzprime ~which is required to survive to
low energies to preserve proton longevity \cite{ps,fm1}.  

In our analysis we vary the unification scale in the range
$2\cdot 10^{16}-5\cdot 10^{17}{\rm GeV}$. The lower scale is the natural 
MSSM unification scale \cite{mssmunification}, $M_X$, whereas the higher scale corresponds to the 
heterotic--string unification scale \cite{kaplunovsky}, $M_S$. 
This factor of 20 discrepancy was discussed in \cite{dienes} and it 
was concluded that intermediate
matter thresholds contributed enough to overcome the difference,
allowing coupling unification in a
wide class of realistic free--fermionic string models \cite{gcu}. 
From the spectra of our models, we will see
that it is natural to include intermediate matter thresholds to achieve string unification. 
It has also been demonstrated that nonperturbative effects arising 
in heterotic M--theory \cite{hw} can push the unification scale down to the MSSM unification 
scale \cite{witten}. Our aim here is to study, qualitatively, the question of gauge coupling unification 
in the LRS heterotic--string models. In particular, to demonstrate that a low scale
$Z^\prime$ in these models is incompatible with the gauge coupling data at the
electroweak scale. The  novel feature of the LRS models  is the 
$U(1)_{Z^\prime}$ charge assignments.  These admit an $E_8$ embedding and therefore
similar charge assignments also arise in heterotic M--theory and so we take the
unification scale to vary between $M_X$ and $M_S$ 
to allow for the possible nonperturbative effects. 
We contrast the analysis in the LRS heterotic--string models 
with the models that admit the $E_6$ embedding of the 
$U(1)_{Z^\prime}$ charges.  In both models there are four intermediate scales
between $M_S$ and $M_Z$, corresponding to:
\begin{itemize}
\item[$M_R$:] \textit{$SU(2)_R$ breaking scale.}  The neutral components of
${\mathcal{H}}_R + {\bar{\mathcal{H}}}_R$ acquire a VEV to break the $SU(2)_R$ symmetry
and leave the \Uzprime ~unbroken.
\item[$M_D$:] \textit{Colour triplet scale.}  
The additional colour triplets in our model acquire a mass at this scale. 
This will also resolve the discrepancy between the MSSM unification 
scale and string scale unification.
\item[$M_{Z^\prime}$:]\textit{$U(1)_{Z^\prime}$ ~breaking scale.}
The $U(1)_{Z^\prime}$ is broken at this scale by singlets acquiring VEVs. 
The anomaly cancelling doublets also acquire mass at this scale and only
the MSSM spectrum survives to lower scales.
\item[$M_{\mbox{\tiny{SUSY}}}$:] \textit{Supersymmetry breaking scale.}
The current bounds from the LHC will be included here to get a
phenomenologically viable supersymmetry scale. Only the SM states remain down
to the $M_Z$--scale, at which the gauge data is extracted.
Threshold corrections for the top quark and Higgs boson are 
included in the analysis.    
\end{itemize}
In addition, due to the extra abelian gauge symmetry acting as our
proton protector, $M_{Z^\prime}$ should be sufficiently low in order for
adequate suppression of induced PDMOs \cite{ps,fm1} . 
By starting from the string scale and evolving the couplings down to $M_Z$,
our analysis may test whether the predictions of these models are in
accordance with low--energy experimental data.  

\subsubsection*{Low--energy inputs}
For our analysis, we take the following values for the masses and 
couplings \cite{pdg}:
\begin{align}
\bes
M_Z &= 91.1876\pm0.0021 \mbox{ GeV}\\
\alpha^{-1}&\equiv\alpha_{\mbox{\tiny{e.m.}}}^{-1}\left(M_Z\right)=127.944\pm0.014
\ees
&&
\bes
\left.\sin^2\theta_W\left(M_Z\right)\right|_{\overline{\mbox{\tiny{MS}}}}
&=0.23116\pm0.00012\\
\alpha_3\left(M_Z\right)&=0.1184\pm0.0007.
\ees
\label{pdg}
\end{align}
We also include the top quark mass of $M_t\sim 173.5$ GeV \cite{pdg}
and the Higgs boson mass of $M_H\sim 125$ GeV \cite{higgs} in our analysis.  
\subsubsection*{Renormalization Group Equations}
For the analyses of both models, we follow \cite{dienes}.
String unification implies that the SM gauge couplings are 
unified at the heterotic--string scale. 
The one--loop renormalization group equations (RGEs)
for the couplings are given by
\beq \label{rge}
\frac{4\pi}{\alpha_i\left(\mu\right)}=k_i\frac{4\pi}
{\alpha_{\mbox{\tiny{string}}}}+
\beta_i\log\frac{M^2_{\mbox{\tiny{string}}}}{\mu^2}+
\Delta_i^{\left(\mbox{\tiny{total}}\right)},
\eeq
where $\beta_i$ are the one--loop beta--function coefficients, 
and $\Delta_i^{\left(\mbox{\tiny{total}}\right)}$ represents possible
corrections from the additional gauge or matter states.
By solving the one--loop RGEs
we obtain expressions for $\sin^2\theta_W\left(M_Z\right)$ and
$\alpha_3\left(M_Z\right)$.
In each model, we initially assume the MSSM spectrum between the
string scale, $M_S$,
and the $Z$ scale, $M_Z$, and treat all perturbations as effective
correction terms.
At the string unification scale we have
\beq
\alpha_S\equiv\alpha_3(M_S)=\alpha_2(M_S)=k_1\alpha_Y(M_S),
\eeq
where $k_1={5}/{3}$ is the canonical $SO(10)$ normalisation.
Thus, the expression for 
$\left.\sin^2\theta_W\left(M_Z\right)\right|_{\overline{MS}}$
takes the general form \cite{dienes}
\begin{align}
\left.\sin^2\theta_W\left(M_Z\right)\right|_{\overline{MS}} = 
\Delta_{\mbox{\tiny {MSSM}}}^{\sin^2\theta_W}+
\Delta_{\mbox{\tiny {I.M.}}}^{\sin^2\theta_W}+
\Delta_{\mbox{\tiny {L.S.}}}^{\sin^2\theta_W}+
\Delta_{\mbox{\tiny {I.G.}}}^{\sin^2\theta_W}+
\Delta_{\mbox{\tiny {T.C.}}}^{\sin^2\theta_W}
\end{align}
with $\left.\alpha_3\left(M_Z\right)\right|_{\tiny{\overline{MS}}}$ 
taking similar form with corresponding $\Delta^{\alpha_3}$ corrections.
Here $\Delta_{\rm MSSM}$ represents the one--loop
contributions from the spectrum of the MSSM 
between the unification scale and the $Z$ scale. 
The following three $\Delta$ terms correspond to
corrections from the intermediate matter thresholds,
the light SUSY thresholds, and the intermediate 
vector bosons corresponding to the $SU(2)_R$ symmetry 
breaking. The last term,
\beq
\Delta_{\mbox{\tiny {T.C.}}}^{\sin^2\theta_W} = 
\Delta_{\mbox{\tiny {H.S.}}}^{\sin^2\theta_W}+
\Delta_{\mbox{\tiny {Yuk.}}}^{\sin^2\theta_W}+
\Delta_{\mbox{\tiny {2-loop}}}^{\sin^2\theta_W}+
\Delta_{\mbox{\tiny {Conv.}}}^{\sin^2\theta_W},
\eeq 
includes the corrections due to heavy string thresholds, and those arising
from Yukawa couplings, two--loops and scheme conversion. These corrections
are small and are neglected for this demonstrative analysis.

For $\sin^2\theta_W\left(M_Z\right)$ we obtain
\begin{align}
\bes \label{s2w}
\Delta_{\mbox{\tiny{MSSM}}}^{\sin^2\theta_W}&=
\frac{1}{1+k_1}\left[1-\frac{\alpha}{2\pi}
\left(11-k_1\right)\log\frac{M_S}{M_Z}\right];\\
\Delta_{\mbox{\tiny{I.M.}}}^{\sin^2\theta_W}&=
\frac{1}{2\pi}
\sum_i
\frac{k_1\alpha}{\left(1+k_1\right)}
\left(\vphantom{\frac{1}{2}}\beta_{2_i}-\beta_{1_i}\right)
\log\frac{M_S}{M_{i}};\\
\Delta_{\mbox{\tiny{L.S.}}}^{\sin^2\theta_W}&=
\frac{1}{2\pi}\frac{k_1\alpha}{\left(1+k_1\right)}
\left(\beta_{1_{\mbox{\tiny{L.S.}}}}-
\beta_{2_{\mbox{\tiny{L.S.}}}}\vphantom{\frac{1}{1}}\right)
\log\frac{M_{\mbox{\tiny{SUSY}}}}{M_Z},
\ees
\end{align}
where $\alpha=\alpha_{\mbox{\tiny{e.m.}}}\left(M_Z\right)$ and $M_i$
are the intermediate
gauge and matter scales discussed earlier.
Similarly for $\alpha_3\left(M_Z\right)$, we have:
\begin{align}
\bes \label{a3}
&\Delta_{\mbox{\tiny{MSSM}}}^{\alpha_3}=\frac{1}{1+k_1}
\left[\frac{1}{\alpha}-
\frac{1}{2\pi}\left(\vphantom{\frac{1}{1}}15+
3 k_1\right)\log\frac{M_S}{M_Z}\right];\\
&\Delta_{\mbox{\tiny{I.M.}}}^{\alpha_3}=
\frac{1}{2\pi}\frac{1}{\left(1+k_1\right)}
\sum_i\left[\left(1+k_1\right)\beta_{3_i}-\left(\beta_{2_i}+
k_1\beta_{1_i}\right)\vphantom{\frac{1}{\alpha}}\right]\log\frac{M_S}{M_{i}};\\
&\Delta_{\mbox{\tiny{L.S.}}}^{\alpha_3}=
-\frac{1}{2\pi}\frac{1}{\left(1+k_1\right)}
\left[\left(1+k_1\right)\beta_{3_{\mbox{\tiny{L.S.}}}}-
\left( \beta_{2_{\mbox{\tiny{L.S.}}}}+
k_1\beta_{1_{\mbox{\tiny{L.S.}}}}\right)\vphantom{\frac{1}{\alpha}}\right]
\log\frac{M_{\mbox{\tiny{SUSY}}}}{M_Z}.\\
\ees
\end{align}
A subtle issue in the analysis of gauge coupling unification
in string models is
the normalisation of the $U(1)$ generators. In GUTs the 
normalisation of abelian generators is fixed by their embedding in
non--abelian groups. However, in string theory the non--abelian symmetry 
is not manifest, and the proper normalisation of the $U(1)$ currents 
is obscured. The $U(1)$ normalisation in string models that 
utilise a world--sheet conformal field theory construction is
fixed by their contribution to the conformal dimensions of 
physical states. The procedure for fixing the normalisation
was outlined in \cite{dienes, Dienes1996}
and we repeat it here for completeness.

In the free fermionic heterotic--string models,
the Ka\v{c}--Moody level of non--abelian
group factors is always one. In general, a given $U(1)$ current, $U$, in the 
Cartan subalgebra of the four dimensional gauge group, is a combination
of the simple world--sheet currents 
$U(1)_{f}\equiv f^*f$, corresponding to individual world--sheet fermions,
$f$.  $U$ then takes the form
$U=\sum_fa_f~U(1)_f$, where the $a_f$ are model dependent coefficients.
Each $U(1)_f$ is normalised to one, so that
$\braket{U(1)_f,U(1)_f}=1$, and each of the linear combinations
must also be normalised to one.
The proper normalisation coefficient for the linear combination $U$
is given by $N=(\sum_fa_f^2)^{-\frac{1}{2}}$, and the properly normalised
$U(1)$ current is, thus, given by
$\hat{U}(1)=N\cdot U$.
 
In general, 
the Ka\v{c}--Moody level, $k$, of a $U(1)$ generator can be
deduced from the operator
product expansion between two of the $U(1)$ currents, and is given by
\beq \label{KacMoodynorm}
k= 2 N^{-2} = 2\sum_fa_f^2.
\eeq
The result is generalised to $k=\sum_ia_i^2k_i~$ when the $U(1)$ 
is a combination 
of several $U(1)$s with different normalisations.
This procedure is used to determine the Ka\v{c}--Moody level,
$k_1$, of the weak--hypercharge 
generator, as well as that of any other $U(1)$ combination
in the effective low--energy field theory.

In the LRS heterotic--string models, the $SO(10)$ symmetry is broken
to $SU(3)_C\times U(1)_C \times SU(2)_L\times SU(2)_R$, where the
combinations of world--sheet currents 
\begin{align}
\frac{1}{3}
\left(\bar{\psi}^*_1\bar{\psi}_1+
\bar{\psi}^*_2\bar{\psi}_2+\bar{\psi}^*_3\bar{\psi}_3\right)
\end{align}
and
\begin{align}
\frac{1}{2}\left(\bar{\psi}^*_4\bar{\psi}_4+\bar{\psi}^*_5\bar{\psi}_5\right)
\end{align}
generate $U(1)_C$ and $T_{3_R}$, respectively, where the latter is the diagonal 
generator of $SU(2)_R$. The weak--hypercharge is then given by 
\begin{align}
\bes
U(1)_Y=T_{3_R}+\frac{1}{3}U(1)_C. 
\ees
\end{align}
The symmetry of $SU(2)_R$ is incorporated in the analysis at the $M_R$
scale, where  above this scale the multiplets are in representations 
of the LRS gauge group and below the $M_R$ scale they are in 
SM representations. The weak-hypercharge coupling relation is given by

\beq\label{hyperembed}
\frac{1}{\alpha_1(M_R)}=\frac{1}{\alpha_{2R}(M_R)}+
\frac{k_C}{9}\frac{1}{\alpha_{\hat{C}}(M_R)}
= \frac{1}{\alpha_{2R}(M_R)}+
\frac{2}{3}\frac{1}{\alpha_{\hat{C}}(M_R).}
\eeq
Here we have used (\ref{KacMoodynorm}) to find that the 
Ka\v{c}--Moody level of $U(1)_C$ is $k_C=6$. 
Again using (\ref{KacMoodynorm}) we find that $k_1=\frac{5}{3}$ as expected.
This reproduces the expected result at the unification scale
\begin{align}
\sin^2\theta_W\left(M_S\right) =
\frac{1}{1+k_1}\equiv\frac{3}{8}.
\end{align}

\subsection{Coupling unification in LRS heterotic--string models}
This model is an example of a three generation, free
fermionic model that yields an unbroken, anomaly free
$U(1)$ symmetry. Heterotic--string models with this
property break the $SO(10)$ symmetry to the
left--right symmetric subgroup \cite{lrs} and are therefore supersymmetric and
completely free of gauge and gravitational
anomalies. The $U(1)_\zeta$ symmetry
in the string models
is an anomaly free, family universal symmetry that forbids the dimension four,
five and six PDMOs, while allowing for the SM fermion mass terms.
A combination of
$U(1)_\zeta$, $U(1)_{B-L}$ and $U(1)_{T_{3_R}}$ remains unbroken down to
low energies
and forbids baryon number violation while allowing for lepton number violation.
Hence, it allows for the
generation of small left--handed neutrino masses via a seesaw mechanism,
specifically an extended
seesaw with the singlets,
$\phi$ \cite{lrs,fm1}.  
Proton decay mediating operators are only generated when the $U(1)_{Z^\prime}$
is broken.  Thus, the scale of the $U(1)_{Z^\prime}$ breaking is
constrained by proton 
lifetime limits and can be within reach of the contemporary experiments. 
A field theory model demonstrating these properties was
presented in \cite{fm1}.  

\subsubsection*{Spectrum}

\begin{table}[!h]
\noindent 
{\small
\begin{center}
{\tabulinesep=1.2mm
\begin{tabu}{|l|ccc|c|c|c|c|c|}
\hline
Field &$\hphantom{\times}SU(3)_C$&$\times SU(2)_L $&
$\times SU(2)_R$&${U(1)}_C$&${U(1)}_\zeta$&$\beta_3$&$\beta_{2L}$& $\beta_Y$\\
%\hline
\hline
$Q_L^i$&$3$&$2$&$1$&$+\frac{1}{2}$&$-\frac{1}{2}$&$1$&
$\frac{3}{2}$&$\frac{1}{6}$\\
$Q_R^i$&$\bar{3}$&$ 1$&$ 2$&$-\frac{1}{2}$&$+\frac{1}{2}$
&$1$&$0$&$\frac{5}{3}$\\
$L_L^i$&$1$&$2$&$1$&$-\frac{3}{2}$&$-\frac{1}{2}$&$0$ 
&$\frac{1}{2}$&$\frac{1}{2}$\\
$L_R^i$&$1$& $1$&$ 2$&$+\frac{3}{2}$&$+\frac{1}{2}$&$0 $ &$0 $&$1$\\
\hline
$H_0$&$1$&$ 2$& $2$&$\hphantom{+}0$&$\hphantom{+}0$&$0 $ &$1 $&$1$\\
\hline
$H_{L}^{ij}$&$1$&$2$&$1$&$+\frac{3}{2}$&$+\frac{1}{2}$&$0 $ 
&$\frac{1}{2} $&$\frac{1}{2} $\\
$H_{L}^{\prime \hphantom{\prime} ij}$&$1$&$2$&$1$&$-\frac{3}{2}$
&$+\frac{1}{2}$&$0 $ &$\frac{1}{2} $&$\frac{1}{2} $\\
$H_{R}^{ij}$&$1$&$1$&$2$&$-\frac{3}{2}$&$-\frac{1}{2}$&$0 $ &$0 $&$1 $\\
$H_{R}^{\prime \hphantom{\prime}ij}$&$1$&$1$&$2$&$+\frac{3}{2}$
&$-\frac{1}{2}$&$0$&$0 $&$1 $\\
\hline
$D^n$&$3$&$1$&$1$&$+1$&$\hphantom{+}0$&$\frac{1}{2} $ 
&$0 $&$\frac{1}{3} $\\
$\bar{D}^n$&$\bar{3}$&$1$&$1$&$-1$&$\hphantom{+}0$
&$\frac{1}{2} $ &$0 $&$\frac{1}{3} $\\
\hline
${\cal H}_R$&$1$&$1$&$2$&$+\frac{3}{2}$&$-\frac{1}{2}$&$0 $ 
&$\frac{3}{5} $&$1$\\
$\bar{\cal H}_R$&$1$&$1$&$2$&$-\frac{3}{2}$&$+\frac{1}{2}$&$0 $ 
&$\frac{3}{5} $&$1$\\
\hline
$S^i$&$1$&$1$&$1$&$\hphantom{+}0$&$-1$&$0$ &$0$&$0$\\
$\bar{S}^i$&$1$&$1$&$1$&$\hphantom{+}0$&$+1$&$0$ &$0$&$0$\\
$\phi^a$&$1$&$1$&$1$&$\hphantom{+}0$&$\hphantom{-}0$&$0$ &$0 $&$0$\\
\hline
\end{tabu}}\end{center}
}
\caption{\label{table1}\it
High scale spectrum and
$SU(3)_C\times SU(2)_L\times SU(2)_R\times U(1)_C\times U(1)_E$ 
quantum numbers, with $i=1,2,3$ for the three light 
generations, $j=1,2$ for the number of doublets required by 
anomaly cancellation, $n=1,...,k$, and $a=1,...,p$.  The $\beta_i$ show 
the contributions for each state, relevant for the RGE analysis later.}
\end{table}

The spectrum of our model above the left--right symmetry breaking scale is 
summarised in Table \ref{table1}. The spectrum below the intermediate symmetry 
breaking scale is shown in Table \ref{table3}. The spectra above and below the 
$SU(2)_R$ breaking scale are both free of all gauge and gravitational anomalies. 
Hence, the $U(1)_{Z^\prime}$ combination given in equation
(\ref{U1zprime}) is viable to 
low energies.

\begin{table}[!h]
\noindent 
{\small
\begin{center}
{\tabulinesep=1.2mm
\begin{tabu}{|l|cc|c|c|c|c|c|c|}
\hline
Field&$\hphantom{\times}SU(3)_C$&$\times SU(2)_L $
&$T_{3R}$&${U(1)}_Y$&${U(1)}_{Z^{\prime}}$&$\beta_3$&$\beta_{2L}$& $\beta_Y$\\
%\hline
\hline
$Q_L^i$&$3$&$2$&$\hphantom{+}0$&$+\frac{1}{6}$
&$-\frac{2}{5}$&$1$&$\frac{3}{2}$&$\frac{1}{6}$\\
$u_L^{c\hphantom{\prime}i}$&$\bar{3}$&$1$&$-\frac{1}{2}$
&$-\frac{2}{3}$&$+\frac{3}{5}$&$\frac{1}{2}$&$0$&$\frac{4}{3}$\\
$d_L^{c\hphantom{\prime}i}$&$\bar{3}$&$1$&$+\frac{1}{2}$
&$+\frac{1}{3}$&$+\frac{1}{5}$&$\frac{1}{2}$&$0$&$\frac{1}{3}$\\
$L_L^i$&$1$&$2$&$\hphantom{+}0$&$-\frac{1}{2}$
&$-\frac{4}{5}$&$0$&$\frac{1}{2}$&$\frac{1}{2}$\\
$e_L^{c\hphantom{\prime}i}$&$1$&$1$&$-\frac{1}{2}$
&$+1$&$+\frac{3}{5}$&$0$&$0$&$1$\\
$\nu_L^{c\hphantom{\prime}i}$&$1$&$1$&$+\frac{1}{2}$
&$\hphantom{+}0$&$+1$&$0$&$0$&$0$\\
\hline
$H^u$&$1$&$2$&$+\frac{1}{2}$&$+\frac{1}{2}$
&$-\frac{1}{5}$&$0$&$\frac{1}{2}$&$\frac{1}{2}$\\
$H^d$&$1$&$2$&$-\frac{1}{2}$&$-\frac{1}{2}$
&$+\frac{1}{5}$&$0$&$\frac{1}{2}$&$\frac{1}{2}$\\
\hline
$H^i_{L}$&$1$&$2$&$\hphantom{+}0$&$+\frac{1}{2}$
&$+\frac{4}{5}$&$0$&$\frac{3}{2}$&$\frac{3}{2}$\\
$H^{\prime\hphantom{\prime}i}_{L}$&$1$&$2$&$\hphantom{+}0$
&$-\frac{1}{2}$&$+\frac{1}{5}$&$0$&$\frac{3}{2}$&$\frac{3}{2}$\\
$E^{i}_{R}$&$1$&$1$&$-\frac{1}{2}$&$-1$&$-\frac{3}{5}$&$0$&$0$&$1$\\
$N^i_{R}$&$1$&$1$&$+\frac{1}{2}$&$\hphantom{+}0$&$-1$&$0$&$0$&$0$\\
$E^{\prime\hphantom{\prime}i}_{R}$&$1$&$1$&$+\frac{1}{2}$&$+1$
&$-\frac{2}{5}$&$0$&$0$&$1$\\
$N^{\prime\hphantom{\prime}i}_{R}$&$1$&$1$&$-\frac{1}{2}$
&$\hphantom{+}0$&$\hphantom{+}0$&$0$&$0$&$0$\\
\hline
$D^n$&$3$&$1$&$\hphantom{+}0$&$+\frac{1}{3}$
&$+\frac{1}{5}$&$\frac{1}{2}$&$0$&$\frac{1}{3}$\\
$\bar{D}^n$&$\bar{3}$&$1$&$\hphantom{+}0$
&$-\frac{1}{3}$&$-\frac{1}{5}$&$\frac{1}{2}$&$0$&$\frac{1}{3}$\\
\hline

$S^i$&$1$&$1$&$\hphantom{+}0$&$\hphantom{+}0$&$-1$&$0$&$0$&$0$\\
$\bar{S}^i$&$1$&$1$&$\hphantom{+}0$&$\hphantom{+}0$&$+1$&$0$&$0$&$0$\\
$\phi^a$&$1$&$1$&$\hphantom{+}0$&$\hphantom{-}0$
&$\hphantom{-}0$&$0$&$0$&$0$\\
\hline
\end{tabu}}\end{center}
}
\caption{\label{table3}\it
Low scale matter spectrum and
$SU(3)_C\times SU(2)_L\times U(1)_Y\times U(1)_{Z^{\prime}}$
quantum numbers with $\beta_i$ contributions. }
\end{table}
The heavy Higgs', ${\mathcal H}^k_R+{\bar {\mathcal H}}^k_R$
that break the $SU(2)_R \times U(1)_C\to U(1)_Y$, along a flat direction,
leave the orthogonal combination
\beq
U(1)_{Z^\prime}={1\over5}U_C-{2\over5}T_{3_R}+U_\zeta \label{U1zprime}
\eeq
unbroken. Here, the index $k$ allows for the possibility that the 
heavy Higgs sector contains more than two fields, as is typically the case in
the string constructions.
Further discussion of this model,
including a trilinear level superpotential, can be found in \cite{fm1}.  
Here we notice that the incomplete representations added to the MSSM may cause
problems with gauge coupling unification. The induced gauge
anomalies in the $SU(2)_{L/R}^2\times U(1)_\zeta$ diagrams require the addition of
$H_{L}^{ij}$, 
$H_{L}^{\prime \hphantom{\prime} ij}$
$H_{R}^{ij}$, 
$H_{R}^{\prime \hphantom{\prime}ij}$, which differ from the $E_6$ case. 
The addition of triplets may help subdue any adverse effects and will
also give scope for the inclusion of intermediate matter scales.    

\subsubsection*{Renormalization group analysis}
\begin{figure}\begin{center}
\includegraphics[scale=1]{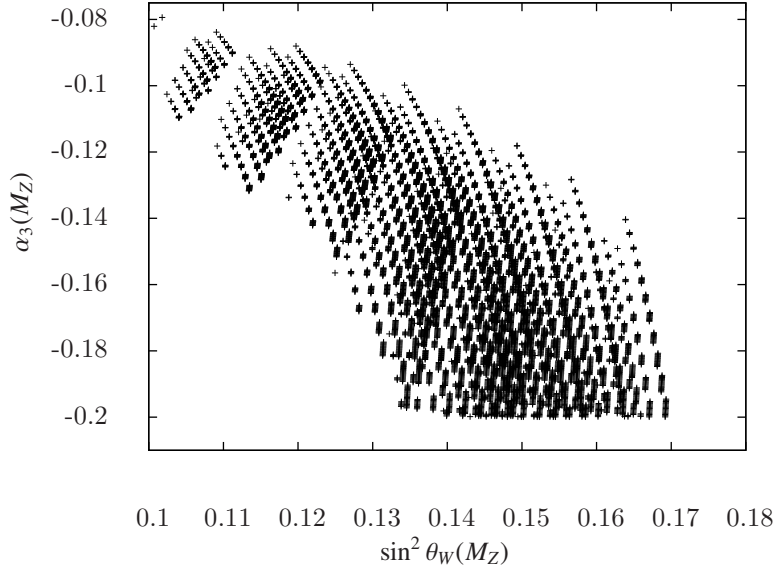} 
\caption{Freely running $\sin^2\theta_W(M_Z)$ and
$\alpha_3(M_Z)$: $\sin^2\theta_W(M_Z)$ vs.
$\alpha_3(M_Z)$ with 
$0.05\lesssim\alpha_{\mbox{\tiny{string}}}\lesssim 0.1$.}
\label{s2wa3SO10}\end{center}
\end{figure}

The properly normalised $\beta$--function coefficients
are shown in Tables \ref{table1}
and \ref{table3}.  The numerical output of equation
(\ref{s2w}) and (\ref{a3}) is generated subject
to the variation of the scales and is displayed in Figure \ref{s2wa3SO10}. 
The intermediate scales are varied to 
find phenomenologically viable areas of
the parameter space. The scales and ranges of $\sin^2\theta_W(M_Z)$ 
and $\alpha_3(M_Z)$  were
first restricted to the experimentally allowed regions and then also allowed to
take values outside this range.
The hierarchy of scales was constrained to be
\beq
M_S \gtrsim M_R > M_D \gtrsim M_{Z^\prime} \gtrsim M_{SUSY} > M_Z.
\eeq
To this end, we restricted the allowed range of
$\sin^2\theta_W(M_Z)$ and $\alpha_3(M_Z)$
 to five sigma deviations from the central values shown in eq (\ref{pdg}). 
The RGEs were run in Mathematica. Restricting the output to the experimentally
constrained interval produced no phenomenologically viable results.
Allowing the values of $\sin^2\theta_W(M_Z)$ and $\alpha_3(M_Z)$
to run freely and restricting the relevant mass scales to (in GeV)
\begin{align}
\bes
2\cdot 10^{16}\leq~ &M_S ~\leq 5\cdot 10^{17};\\
10^9\leq~ &M_R~\leq 5\cdot 10^{17};
\ees
&&
\bes
10^5\leq~&M_D~\leq 10^{12};\\
10^{3}\leq~&M_{Z^\prime},M_{\mbox{\tiny{SUSY}}}~\leq 10^{10},
\ees
\end{align}
also produced no phenomenologically viable results,
as shown in Figure \ref{s2wa3SO10}.

\subsubsection*{Contrasting analysis with $E_6$ embedding of $U(1)_\zeta$ }
\begin{figure}\begin{center}
\includegraphics[scale=1]{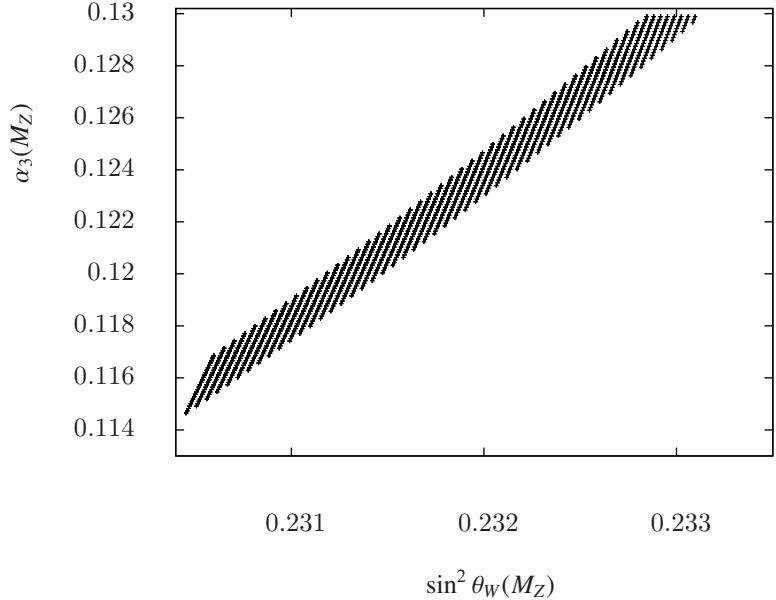} 
\caption{Freely running $\sin^2\theta_W(M_Z)$ and $\alpha_3(M_Z)$:
$\sin^2\theta_W(M_Z)$ vs. $\alpha_3(M_Z)$ with
$0.05\lesssim\alpha_{\mbox{\tiny{string}}}\lesssim 0.1$
for Model II.}\label{E6}\end{center}
\end{figure}

To further elucidate the constraints on the LRS heterotic--string models
arising from coupling unification, we contrast the 
outcome with the corresponding results
when the $U(1)_\zeta$ charges are embedded in $E_6$ representations. 
For models that allow the $E_6$ embedding of the $U(1)_{Z^\prime}$ charges,
the spectrum consists of three
generations of $\mathbf{27}$s that decompose under $SO(10)$ as:
\beq
{\mathbf{27}}^i\to {\mathbf{16}}^i_{\frac{1}{2}}+
{\mathbf{10}}^i_{-1}+{\mathbf{1}}^i_{2}.
\eeq
Under \highgg, ~this results in a similar spectrum to the LRS model. 
The \textbf{16} decomposes exactly as for the LRS model,
\begin{align}
\bes \label{16}
Q_L^i& \sim \left(3,2,1,+\frac{1}{2},+\frac{1}{2}\right); \\
Q_R^i& \sim \left(\bar{3},1,2,-\frac{1}{2},+\frac{1}{2}\right);
\ees
&&
\bes
L_L^i& \sim \left(1,2,1,-\frac{3}{2},+\frac{1}{2}\right); \\ 
L_R^i& \sim \left(1,1,2,+\frac{3}{2},+\frac{1}{2}\right),
\ees
\end{align}
with the proviso that the charges under \Uzeta ~take the same sign.
The \textbf{10} decomposes as
\begin{align} \bes \label{10}
H^i&\sim\left(1,2,2,0,-1\right);
\ees
&&
\bes
D^i&\sim \left(3,1,1,+1,-1\right);
\ees
&&
\bes
\bar{D}^i&\sim \left(\bar{3},1,1,-1,-1\right).
\ees
\end{align}
The remaining singlets are neutral under the SM gauge group
and are used to break the \Uzprime.
In addition to the complete $SO(10)$ representations above,
the $E_6$ spectrum includes a bidoublet, 
\beq
H_0 \sim \left(1,2,2,0,-1\right),
\eeq
that facilitates gauge coupling unification. 
The model also contains the pair of heavy Higgs right--handed doublets, 
\beqa
{\mathcal H}_R+{\bar {\mathcal H}}_R=\left(1 ,1,2,\frac{3}{2}, \frac{1}{2}\right)
+ \left(1 ,1,2,-\frac{3}{2},  -\frac{1}{2}\right), 
\eeqa
that break the intermediate $SU(2)_R$ symmetry.  
We run the RGEs in exactly the same way as shown for the LRS model,
constraining the mass scales to the hierarchy
\beq
M_S \gtrsim M_R \gtrsim M_D = M_{Z^\prime} \gtrsim M_{SUSY} \gg M_Z.
\eeq
In this model we find that unification does occur,
as found in previous literature. 
We note that the phenomenologically viable results 
(see Figure \ref{E6}) required $M_S\sim M_X \sim 2\cdot 10^{16}$~GeV as expected. 
The intermediate scales were found to be (in GeV)
\begin{align}
\bes
1\cdot 10^{13} \leq M_R\leq 1\cdot 10^{16};
\ees
&&
\bes
1\cdot 10^3\leq~&M_D~\leq 1\cdot 10^8;
\ees
&&
\bes
1\cdot10^{3}\leq~&M_{\mbox{\tiny{SUSY}}}~\leq 1\cdot 10^6,
\ees
\end{align}
with $M_{Z^\prime}$ between $1-10^5$~TeV.
In this case we have taken the mass of the vector--like doublets,
$M_{Z^\prime}$, and triplets, $M_D$
to be degenerate, which is the case in $E_6$ inspired models,
as they are generated by the 
same singlet VEV. String models afford more flexibility
that we do not make use of 
in our analysis here. 
Fine--tuning the $M_{\mbox{\tiny{SUSY}}}$ allows for $M_{Z^\prime}$
to be in agreement with current experimental bounds. 

The contrast between the two cases can be elucidated further by examining more 
closely the contributions of the intermediate gauge and matter thresholds to
$\sin^2\theta_W(M_Z)$ and $\alpha_3(M_Z)$.
Using the general expressions in equations (\ref{s2w}) and
(\ref{a3}) we find that, 
in the case of the spectrum and charge assignments in the
LRS heterotic--string model, 
shown in Tables \ref{table1} and \ref{table3},
the threshold corrections from intermediate gauge 
and matter scales are given by 
\begin{align}
\bes\label{deltalrs}
\delta\left(\sin^2\theta_W(M_Z)\right)_{\mbox{\tiny{I.T.}}} & =
\frac{1}{2\pi}\frac{k_1\alpha}{1+k_1}
\left(
\frac{12}{5}\log\frac{M_S}{M_R}-
\frac{24}{5}\log\frac{M_S}{M_{Z^\prime}}-
\frac{2n_D}{5}\log\frac{M_S}{M_D}\right), \\
\delta\left(\alpha_3(M_Z)\right)_{\mbox{\tiny{I.T.}}} & =
\frac{1}{2\pi}
\left(
\frac{3}{2}\log\frac{M_S}{M_R}-
9\log\frac{M_S}{M_{Z^\prime}}+
\frac{3n_D}{4}\log\frac{M_S}{M_D}\right).
\ees
\end{align}
In the case of models that admit an $E_6$ embedding of the charges,
the same threshold corrections are given by 
\begin{align}
\bes\label{deltae6e}
\delta\left(\sin^2\theta_W(M_Z)\right)_{\mbox{\tiny{I.T.}}}  &=
\frac{1}{2\pi}\frac{k_1\alpha}{1+k_1}
\left(
\frac{12}{5}\log\frac{M_S}{M_R}+
\frac{6}{5}\log\frac{M_S}{M_H}-
\frac{6}{5}\log\frac{M_S}{M_D}\right),\\
\delta\left(\alpha_3(M_Z)\right)_{\mbox{\tiny{I.T.}}}& =
\frac{1}{2\pi}
\left(
\frac{3}{2}\log\frac{M_S}{M_R}-
\frac{9}{4}\log\frac{M_S}{M_H}+
\frac{9}{4}\log\frac{M_S}{M_D}\right).
\ees
\end{align}
If we take $M_S$ to coincide with the MSSM unification 
scale and with $M_R$ as well, then the first lines in equations
(\ref{s2w}) and (\ref{a3}), which only contain 
the MSSM contributions, are in good agreement with the observable data.
The corrections arising from the intermediate
gauge and matter thresholds in equations (\ref{deltalrs})
and (\ref{deltae6e}) then have to cancel. 
We see from 
equation (\ref{deltalrs}) that the corrections from the
intermediate doublet and triplet thresholds 
contribute with equal sign in $\sin^2\theta_W(M_Z)$.
For $\alpha_3(M_Z)$, the corrections
from these thresholds contribute with opposite sign,
but the contribution of the doublets outweigh
the contribution of the triplets. We may compensate for
the negative contribution from
the extra doublets by lowering the $SU(2)_R$
breaking scale. Requiring that
$m_{\nu_\tau}\lesssim 1{\rm eV}$ necessitates that
$M_R\geq 10^9{\rm GeV}$.
Keeping the extra triplets at the GUT scale, and the
$Z^\prime$ scale at $10^{12}{\rm GeV}$
then yields rough agreement with $\sin^2\theta_W(M_Z)$
but gross disagreement with
$\alpha_3(M_Z)$. Lowering the triplet scale improves the agreement with 
$\alpha_3(M_Z)$ but conflicts with the data for $\sin^2\theta_W(M_Z)$. 
We therefore conclude that a low scale $Z^\prime$ in the
LRS heterotic--string models
is incompatible with the gauge data at the $Z$--boson scale.  
In contrast, from equation (\ref{deltae6e}) we see that the corresponding
corrections cancel each other, provided that $M_H=M_{Z^\prime}=M_D$.
This is the case as both are
generated by the $Z^\prime$ breaking VEV. This cancellation is,
of course, the well known
cancellation that occurs when the representations fall into $SU(5)$ multiplets. 
Allowing $M_R$ to be at $10^{15}{\rm GeV}$ then compensates for the SUSY 
threshold at $1{\rm TeV}$, enabling accommodations of the low--energy data, as 
illustrated in Figure \ref{E6}.

\section{String models with $E_6$ embedding}

The low scale $Z^\prime$ in the string models is, in essence, a combination 
of the Cartan generators, $U(1)_{1,2,3}$, that are generated by the 
right--moving complex world--sheet fermions ${\bar\eta}^{1,2,3}$,
together with a $U(1)$ symmetry, embedded in the $SO(10)$ 
GUT, and is orthogonal to the weak hypercharge. Whether, or not, 
the symmetry is anomaly free depends on the specific 
symmetry breaking  pattern induced by the GGSO projections. 
As we discussed above, in the FSU5, PS and SLM the symmetry is
anomalous, whereas in the LRS models it is anomaly free. 
The difference stems from the fact that in the former cases 
the combination for $U(1)_\zeta$ admits the $E_6$ embedding but
in the latter it does not. On the other hand, as we have seen 
in Section \ref{gca}, the $E_6$ embedding allows for 
compatibility with the low scale gauge coupling data. 
The ${Z^\prime}$ in the LRS models, which 
do not admit the $E_6$ embedding, is constrained to 
be heavier than at least $10^{12}$GeV. Gauge coupling  data,
therefore, seems to indicate that the $E_6$ embedding 
of the charges is necessary. We emphasize that the 
indication is that the charges must admit an $E_6$ embedding and 
\textit{not} that the $E_6$ symmetry is actually realised.
An illustration of this phenomenon is the existence
of self--dual models under the spinor--vector duality 
without $E_6$ enhancement \cite{spinvecdual}. 
The question then arises as to how one constructs heterotic--string models with
anomaly free $U(1)_\zeta$, which admit an $E_6$ embedding. 
Here we discuss how viable heterotic--string models with $E_6$
embedding of the $U(1)_{Z^\prime}$ charges may be obtained. The main
constraint being that the extra $U(1)$ symmetry has to be anomaly free.  
For this purpose, we first give a general overview as to how the gauge 
symmetry is generated in the string models. 

The vector bosons that generate the four dimensional 
gauge group in the string models arise from two principal sectors: 
the untwisted sector and the 
sector $x=\{{\bar\psi}^{1, \cdots, 5}, {\bar\eta}^{1,2,3}\}$. In the 
$x$--sector the complex right--moving world--sheet fermions, that 
generate the Cartan subalgebra of the observable gauge group, are
all periodic. At the level of the $E_8\times E_8$ heterotic--string
in ten dimensions, the vector bosons of the observable $E_8$ are 
obtained from the untwisted sector and from the $x$--sector. 
Under the decomposition $E_8\rightarrow SO(16)$, the adjoint 
representation decomposes as $\mathbf{248}\rightarrow
\mathbf{120}+\mathbf{128}$, where
the adjoint $\mathbf{120}$ representation is obtained from the untwisted
sector and the spinorial $\mathbf{128}$ representation is obtained 
from the  $x$--sector. The set $\{{\bf 1},S, x, \zeta\}$
produces a model with $N=4$ spacetime supersymmetry
in four dimensions. The gauge symmetry arising in this model,
at a generic point in the compactified space,
is either $E_8\times E_8$ or $SO(16)\times SO(16)$ depending 
on the GGSO phase $c({x\atop\zeta})=\pm1$.

Adding the basis vectors $b_1$ and $b_2$ reduces the spacetime 
supersymmetry to $N=1$. The observable gauge symmetry
reduces from $E_8$ to $ E_6\times U(1)^2$ or
$SO(16)\rightarrow SO(10)\times U(1)^3$. Additional vectors
reduce the gauge symmetry further. Aside from the model
of \cite{bfgrs}, all the quasi--realistic free fermionic 
models follow the second symmetry breaking pattern. That is, 
in all these models, the vector bosons arising from the 
$x$--sector are projected out. 

We consider, then, the symmetry breaking pattern induced by 
the following boundary condition assignments in two 
separate basis vectors
\beqn
&1.&b\{{{\bar\psi}_{1\over2}^{1\cdots5}}\}=
\{{1\over2}{1\over2}{1\over2}{1\over2}
        {1\over2}\}\Rightarrow SU(5)\times U(1),\label{su51breakingbc}\\
&2.&b\{{{\bar\psi}_{1\over2}^{1\cdots5}}\}=\{1~ 1\, 1\, 0\, 0\}\,
  \Rightarrow SO(6)\times SO(4).
\label{so64breakingbc}
\eeqn
The assignment in equation (\ref{su51breakingbc}) reduces the untwisted 
$SO(10)$ gauge symmetry to $SU(5)\times U(1)$, however the 
assignment in eq. (\ref{so64breakingbc}) reduces it to 
$SO(6)\times SO(4)$. Thus, the inclusion of equations
(\ref{su51breakingbc}) and (\ref{so64breakingbc})
in two separate boundary condition basis vectors 
reduces the $SO(10)$ gauge symmetry to 
$SU(3)_C\times SU(2)_L\times U(1)_C\times U(1)_L$, 
where $2 U(1)_C={3}U(1)_{B-L}$ and $U(1)_L=2 U(1)_{T_{3_R}}$. 
For appropriate 
choices of the GGSO projection coefficients, the vector bosons
arising from the $x$--sector enhance the 
$SU(3)\times SU(2)\times U(1)^2\times U(1)_\zeta$ 
arising from the untwisted sector to
$SU(4)_C\times SU(2)_L\times SU(2)_R\times U(1)_{\zeta^\prime}$, where
\beqn
U(1)_4 & = & U(1)_C+3 U(1)_L - 3 U(1)_\zeta; \label{u14}\\
U(1)_2 & = & U(1)_C+ U(1)_L +    U(1)_\zeta; \label{u12}\\
U(1)_{\zeta^\prime} & = & -3 U(1)_C+3 U(1)_L + U(1)_\zeta. \label{u1zetaprime}
\eeqn 
$U(1)_4$ and $U(1)_2$ are embedded in $SU(4)_C$ and $SU(2)_R$,
respectively, and $U(1)_\zeta$ is given by equation (\ref{u1zeta}).
The matter representations charged under this group arise 
from the sectors $b_j$ and are complemented by states from
$b_j+x$ to form the ordinary representations of the Pati--Salam 
model. The difference, as compared to the Pati--Salam string models
of \cite{alr}, is that $U(1)_{\zeta^\prime}$ is anomaly free. 
The reason is that all the states of the $\mathbf{27}$ representation of
$E_6$ are retained in the spectrum, whereas in the Pati--Salam
models of \cite{alr} the corresponding states are projected out. 
The symmetry breaking of the Pati--Salam $SU(4)_C\times SU(2)_R$
group is induced by the VEV of the 
heavy Higgs in the
$({\bar 4}, 1,2)_{-{1\over2}}\oplus ({4}, 1,2)_{+{1\over2}}$
representation of
$SU(4)_C\times SU(2)_L\times SU(2)_R\times U(1)_{\zeta^\prime}$.
In addition to the weak--hypercharge, this VEV leaves
the unbroken combination
\beq
U(1)_{Z^\prime} = 
                  {1\over2} U(1)_{B-L} 
                - {2\over3} U(1)_{T_{3_R}}
                + {5\over3} U(1)_{\zeta^\prime},
\label{u1zprime}
\eeq
which is anomaly free and admits the $E_6$ embedding of the charges. 

\section{Conclusions}

In this paper we examined the gauge coupling unification 
constraints imposed on a low scale $Z^\prime$ arising in 
LRS heterotic--string derived models. 
The existence of a low--scale $Z^\prime$ in these models
guarantees that PDMOs are sufficiently suppressed. 
However, we have shown that the hypothesis of a low scale $Z^\prime$ 
in these models is incompatible with the gauge coupling data
at the electroweak scale. 
We contrasted this result with the corresponding result in string models that 
admit an $E_6$ embedding of the $U(1)$ charges. In the latter case
the possibility of a low scale $Z^\prime$ is viable. We further discussed
how heterotic--string models that admit the $E_6$ embedding may 
be obtained in the free fermionic formulation, though an explicit 
three generation viable model is yet to be constructed. 
Similarly, a more complete analysis of the
phenomenological realisation of this $U(1)$ 
symmetry in heterotic--string models is warranted
and will be reported in future publications. 
We also remark that other $U(1)$ symmetries that
have been proposed in the literature 
to suppress proton decay mediating operators \cite{ zphistory, pati} 
have also been invalidated 
due to neutrino masses and other constraints \cite{ps}. 
The enigma of the proton lifetime in heterotic--string unification
continues to serve as an important guide in the search for viable string vacua.

\section*{Acknowledgements}

AEF thanks Subir Sarkar and 
Theoretical Physics  Department at the University of Oxford for hospitality. 
This work was supported in part by the STFC (PP/D000416/1).

\end{document}